\documentclass[epj,nopacs]{svjour}
\usepackage{graphics}
\usepackage{amsmath}

\usepackage{amssymb}

\usepackage{makeidx}

\begin{document}

\newcommand{\bequ}{\begin{equation}}

\newcommand{\eequ}{\end{equation}}
\newcommand{\sg}{\sigma}
\newcommand{\sig}[1]{{\sigma}^{#1}}
\newcommand{\sigo}[1]{{\sigma}_{0}^{#1}}
\newcommand{\intx}{\int d^3x \,}
\newcommand{\Uw}{U_{\omega}}
\newcommand{\Uwsg}{\Uw(\sg)}
\newcommand{\tmeio}{\tfrac{1}{2}}
\newcommand{\meio}{\frac{1}{2}}%
\title{Q-balls: some analytical results}
\author{F. Paccetti Correia\thanks{e-mail: F.Paccetti@thphys.uni-heidelberg.de} \and M.G. Schmidt\thanks{e-mail: M.G.Schmidt@thphys.uni-heidelberg.de}
%\thanks{\emph{Present address:} Insert the address here if needed}
}                     % Do not remove
\institute{Institut f\" ur Theoretische Physik, Universit\" at Heidelberg, Philosophenweg 16, 69120 Heidelberg, Germany}
\abstract{
Motivated by the renewed interest in the role of Q-balls in cosmological evolution, we present a discussion of the main properties of Q-balls, including some new results.
} %end of abstract
\maketitle
\section{Introduction}
\label{intro}

The existence of the Baryon asymmetry in nature is still an unsolved
puzzle. As there is a general agreement that it can't be explained within the
framework of the Standard Model (SM), it has become a common practice to
investigate \emph{supersymmetric} extensions of the SM. However, in
addition to electroweak baryogenesis, these models allow for other
possible mechanisms for generating the B-asymmetry. One of these is the
\emph{Affleck-Dine} (AD) mechanism~\cite{aff85,din96}, where the baryon
number surplus is produced in the late stages of inflation, being stored
in a condensate of supersym\-metric scalars (e.g. a mixture of
\emph{squarks}). But, as Kusenko and Shaposhnikov pointed
out~\cite{kus98}, this homogeneous condensate is not allways the best
way of packing the baryon number: there are non-ho\-mo\-ge\-neous
states, lumps of baryonic matter, with the same B-number but less
energy, which are known as \emph{Q-balls} (or B-balls). The condensate
is therefore expected to colapse, producing a large number of Q-balls.
The B-number, stored in the Q-balls in this way, survives the
\emph{sphaleron} processes if they decay at temperatures below the weak
scale. As shown in \cite{enq99}, in the context of models with
\emph{gravity-mediated supersymmetry breaking} this happens only for
Q-balls with B$\gtrsim 10^{16-17}$, as it is actually realized in such models.
Since their decay is due to baryon and \emph{neutralino} evaporation from the surface this could explain the similarity between the present baryon density and dark matter density, as the neutralinos are candidates to the lightest SUSY particles and therefore to dark matter. It is also possible to explain this similarity, in a \emph{gauge mediated} SUSY breaking scenario, though with a different mechanism\cite{kus98}. In this case, Q-balls are so large that they should survive to the present, constituting therefore dark matter. Their partial evaporation at high temperatures should be responsible for baryogenesis.

Our intention here is not, however, to investigate the role of Q-balls
in baryogenesis, but to discuss their \emph{general} properties. As
was shown by Coleman, the existence of Q-balls is a general feature of
theories with scalars carrying a conserved U(1)-charge
\cite{col85,lee91,kus97}(e.g. B-number). They can be regarded as bound
states of the scalar particles and appear as stable classical solutions
of the equations of motion, i.e. they are non-topological solitons which
arise due to the \emph{non-linear self-interactions} of the scalar
field.

The present paper (see also ref.\cite{pac00}) is a study of the main general properties of Q-balls. We will start in sec.\ref{sec:2} by explaining \emph{why} one should search for solutions of the eq. of motion of the Q-ball type (i.e.
$\phi=\sg(r)e^{i\omega t}$), and \emph{when} they exist. In the case
that such solutions exist, the next rational move is to investigate
their stability. This is what we do in sec.\ref{sec:3}: We prove a
statement which relates the classical stability of a Q-ball with the
dependence of its charge on the internal frequency $\omega$. In
sec.\ref{sec:4} we discuss the limits of small and high internal
frequency using the \emph{thin-wall} and the \emph{thick-wall}
approximation, respectively. These results are then applied to the case
of a model with broken symmetry. Finally we regard in sec.\ref{sec:5} the issue of the meaning of Q-balls in the quantum theory. We close with a recapitulation and discussion of our results.

\section{Non-topological solitons in scalar theories}
\label{sec:2}

To make our investigation as simple as possible we will neglect the
interactions of the scalar sector with the rest of the world (the other
sectors of the theory), and investigate U(1)-invariant scalar theories,
defined in a general way by the lagrangian density

\begin{equation}
                    {\cal L}=|\partial_{\mu}\phi|^2 - \bar{U}(|\phi|).
\end{equation}
In this case, classical dynamics is described by the equation of motion

\begin{equation}\label{12}
                    \partial_{t}^2\phi - \nabla^2\phi + \bar{U}'(|\phi|)\frac{\phi}{2|\phi|} =0,
\end{equation}
for which the charge

\begin{equation}
                      Q[\phi]=\frac{1}{i} \int d^3x ( \phi^{\ast}\partial_{t}\phi - c.c.)
\end{equation}
is conserved. We require the potential to have a minimum at the origin $\phi=0$, with $\bar{U}(0)=0$, which is the same as stating that there is a sector of scalar particles (which we call mesons), that carry the U(1) charge, and have masses $m^2=\meio \bar{U}^{\prime\prime}(0)$.

The purpose of this section is to show that for certain potentials there
are solutions of the eq. of motion, called Q-balls\cite{col85}, of the
form
\begin{equation}
                {\phi(\bf x,t)} = \frac{\sg(r)}{\sqrt{2}}e^{i\omega t},
\end{equation}
with $\sigma(r)$ being a decreasing function of $r=|{\bf x}|\ge 0$ (a ball). The point of this \emph{ansatz} is, that it has the simplest form, which allows the Q-ball to carry \emph{finite} charge.
The energy and charge of the Q-ball are
\begin{equation}\label{106}
                E=\int d^3x\left[ \frac{1}{2}\omega^2\sig{2}(r) + \frac{1}{2}(\nabla\sig{}(r))^2 + U(\sig{}(r))\right],
\end{equation}

\begin{equation}\label{107}
      Q=\omega \intx \sig{2}(r),
\end{equation}
($U(\sg)\equiv\bar{U}(|\phi|)$) and the equation of motion is now

\begin{equation}\label{um}
 \nabla^2\sigma = U'(\sigma)- \omega^2\sigma.
\end{equation}

Since we want to find minima of the energy in sectors of \emph{fixed charge}, it will be usefull to rewrite the energy functional as follows
\begin{equation}\label{EQ}
               E_{Q}[\sg]=\frac{1}{2}\frac{Q^2}{I[\sg]} + \int d^3x\left[\frac{1}{2}(\nabla\sigma)^2 + U(\sigma) \right],
\end{equation}
with
\begin{equation}
 I[\sg]=\int d^3x \,\sigma^2.
\end{equation}

With $E_{Q}$ recast in this form, we can investigate the stability of any Q-ball solution with respect to perturbations inside the subspace of fixed charge and configurations of the form ${\phi({\bf x},t)} = \sigma({\bf x})e^{i\omega t}/\sqrt{2}$. The handling of general variations around the Q-ball solution will be postponed, because, as we will see in section 3, the only relevant configurations with regard to the stability of the Q-balls are those with the above form.
%
%By making the change ${\sigma \to \sg + \delta\sg}$, and picking up only terms up to second order we get:
%\begin{equation}\label{dois}
%\begin{split}
%                \frac{Q^2}{2I[\sg +\delta\sg]}& =\frac{Q^2}{2}\left[I[\sg]+2\intx \sg\,\delta\sg + \intx (\delta\sg)^2\right]^{-1}={}\\
%                & =\frac{Q^2}{2I} \left[1- \frac{2\intx \sg\,\delta\sg}{I} + 4\frac{(\intx \sg\,\delta\sg)^2}{I^2} - \frac{\intx (\delta\sg)^2}{I} + ...\right]={}\\
%                & =\frac{Q^2}{2I} - \omega^2\intx \sg\,\delta\sg - \frac{1}{2}\omega^2\intx (\delta\sg)^2={}\\
%                & \quad + 2\frac{\omega^2}{I}\left(\intx \delta\sg\right)^2 + O(\delta\sg^3),
%\end{split}
%\end{equation}
%
%and
%\begin{multline}
%                \intx\left[\frac{1}{2}(\nabla(\sg + \delta\sg))^2 + U(\sg + \delta\sg)\right]=\\
%                 =\intx \delta\sg(-\nabla^2\sg + U') + \intx\frac{1}{2}\delta\sg(-\nabla^2 + U'')\delta\sg + O(\delta\sg^3),
%\end{multline}\\[12pt]
%
%where we used $Q={\omega}I$ and Taylor expanded $I^{-1}[\sg + \delta\sg]$.

By Taylor expanding $I^{-1}[\sg + \delta\sg]$, we can write the energy's variation at fixed charge in the form
\bequ\label{cinco}
\begin{split}
                \Delta E_{Q}&=\intx \delta\sg(-\nabla^2\sg + U'-\omega^2\sg)\\
                          & \quad + \intx\frac{1}{2}\delta\sg(-\nabla^2 + U''-\omega^2)\delta\sg\\
                          & \quad + 2\frac{\omega^2}{I}\left(\intx \sg\delta\sg\right)^2 + O(\delta\sg^3).
\end{split}
\end{equation}
Note that we traded $Q$ for $\omega$ using $Q=\omega I$ only after performing the variation. It is helpfull to define the following functionals,
\begin{equation}
       U_{\omega}(\sg) \equiv U(\sg) - \frac{1}{2}\omega^2 \sg^2,
\end{equation}
\bequ\label{115}
       S_{\omega}[\sg] \equiv \intx (\frac{1}{2}(\nabla\sg)^2 + U_{\omega}),
\end{equation}
to recast the energy's variation in a simpler way:
\bequ\label{116}
       \Delta E_{Q}=\Delta S_{\omega} + 2\frac{\omega^2}{I}\left(\intx \sg\delta\sg\right)^2 + O(\delta\sg^3),
\end{equation}
where $\Delta S_{\omega}$ is a variation made while keeping $\omega$ fixed.

We are now in position to make some remarks:
\begin{itemize}
                 \item[(i)] Extrema of the functional $S_{\omega}$ at fixed $\omega$ are extrema of the energy $E_{Q}$ at fixed charge $Q$~\cite{kus97}, as we can see from eqs.\eqref{cinco} and \eqref{116}. They satisfy automaticaly the equation of motion eq.\eqref{um}. Now, if $\{\sg_{\omega}(r)\}$ is a set of solutions of this equation, parameterized by $\omega$, $E(Q)\equiv E_Q[\sg_{\omega}]$ can be seen as a function of only $Q$ and $S(\omega)\equiv S_{\omega}[\sg_{\omega}]$ as a function of $\omega$. These two functions are related through a \emph{Legendre transformation}. To see this note that from eq.\eqref{106} it follows that
\bequ\label{eq:14}
               E(Q)  = S(\omega) + \omega Q,
\end{equation}
which defines a Legendre transformation because
\bequ\begin{split}\label{193}
                  \frac{dS(\omega)}{d\omega} & = \int d^3r\,\frac{\delta S_{\omega}}{\delta\sg_{\omega}}\frac{d\sg_{\omega}}{d\omega} + \frac{\partial S(\omega)}{\partial\omega}\\[12pt]
                            & = 0 - \int d^3r\,\sg_{\omega}^2\frac{\partial}{\partial\omega}\frac{1}{2}\omega^2  = - Q(\omega),
\end{split}\end{equation}
where $Q(\omega)\equiv Q[\sg_{\omega}]$. (These expressions will prove to be usefull in sec.\ref{sec:4}.)\\

                 \item[(ii)] Fortunately the problem of finding the solutions of $\frac{\delta S_{\omega}}{\delta\sg}=0$ is a well investigated one. In fact, it is known that for values of $\omega^2$ within a certain range, the extremum of $S_{\omega}$ with the smallest value of $S_{\omega}$ is a decreasing function of $r=|{\bf x}|$, the so-called \emph{bounce}, which satisfies the boundary conditions $\frac{d\sg}{dr}(0)=0$, $\sg(+\infty)=0$~\cite{col77a,col77b,col78}\footnote{These configurations describe quantum tunneling in a real scalar field theory in (2+1)-dimensions with potential $U_{\omega}(\sg)$.}.\\
                 \item[(iii)] Minima of $E_{Q}$ don't need to be minima of $S_{\omega}$ (see eq.\eqref{116}):
                   \bequ
                 \delta^2 E_{Q}\geq 0 \quad \Rightarrow \quad \delta^2 S_{\omega} \geq - 2\frac{\omega^2}{I}\left(\intx \sg\delta\sg\right)^2.
                   \end{equation}
\begin{figure}
\resizebox{0.47\textwidth}{!}{
  \includegraphics{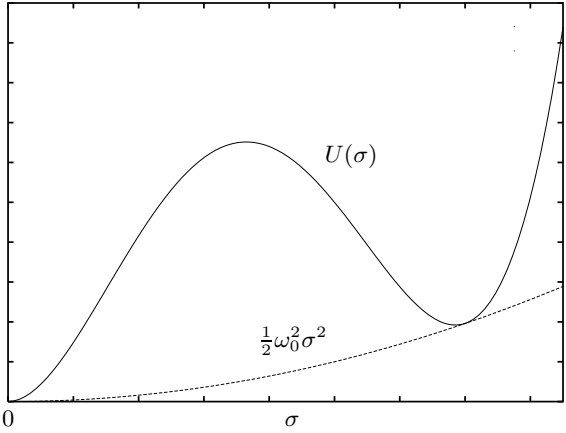}
}
%\vspace{5cm}       % Give the correct figure height in cm
\put(-115,0){$\sigma$}
\put(-222,0){$0$}
%\put(-48,151){\tiny $U(\sg)$}
%\put(-53,142){\tiny $\frac{1}{2}\omega_{0}^2\sg^2$}
\put(-100,100){$U(\sg)$}
\put(-125,30){$\frac{1}{2}\omega_{0}^2\sg^2$}
\caption{The potential $U(\sg)$ supports Q-ball solutions, with squared internal frequency $\omega^2$ in the range $]\omega_0^2,m^2[$, where $\omega_0^2$ is given by eq.\eqref{omegazero}.}
\label{fig:1}       % Give a unique label
\end{figure}
%
%\begin{figure}[]\label{f_um}
%\begin{center}
%\includegraphics[width=8cm]{plot_1.1.eps}
%\put(-45,145){\tiny $U(\sg)$}
%\put(-50,135){\tiny $\frac{1}{2}\omega_{0}^2\sg^2$}
%\end{center}
%\caption{The potential $U(\sg)$ supports Q-ball solutions, with internal frequency $\omega$ in the range $[\omega_0,m]$, where $\omega_0$ is given by eq.\eqref{omegazero}. }
%\end{figure}
%
%
                              This is an important feature, because it is a well known fact in the theory of \emph{bounces} that these solutions are not minima of $S_{\omega}$, but rather saddle points. They have \emph{one} mode ${ \delta\sg_{-1}}$\, with $\frac{\delta^2S_{\omega}}{\delta\sg_{-1}^2}<0$~\cite{col87}.
\end{itemize}

We want now to find the range of $\omega^2$, for which there are configurations of the form discussed in (ii) which satisfy eq.\eqref{um}. For radial-dependent solutions this equation turns to be
\bequ
                 \frac{d^2\sg}{dr^2} = -\frac{2}{r}\frac{d\sg}{dr} + U_{\omega}'(\sg).
\end{equation}
It is easy to recognize this as the equation of motion for a particle of unit mass, with position $\sg$ and time $r$, under the action of the potential $-U_{\omega}(\sg)$ and of a viscous term proportional to the velocity and the inverse of the time. In this mechanical analogon, the boundary conditions are that the particle starts at rest somewhere at the positive axe, and moves towards the origin, which is attained at infinite time. This is only possible if~\cite{col85}
\begin{itemize}
           \item [1)] $\omega^2 \leq m^2$: When the particle approa\-ches the origin the potential behaves like
                 \bequ
                              -U_{\omega}(\sg) \simeq -U_{\omega}(0) - \meio\frac{d^2 U_{\omega}(0)}{d\sg^2}\sg^2 = - \meio(m^2-\omega^2)\sg^2.
                 \end{equation}
                 If $\omega^2 > m^2$, the particle will ultimately start oscillating around the origin as
                 \bequ
                               \sg(r) \sim \frac{\sin{(\sqrt{\omega^2 - m^2}r)}}{r},
                 \end{equation}
                 causing $E_Q[\sg]$ to be infinite.\\
           \item [2)]$\omega^2 > \omega_0^2$, where $\omega_0^2$ is defined as
                 \bequ\label{omegazero}
                               \omega_0^2 \equiv \text{min} \left(\frac{2U(\sg)}{\sg^2}\right)\equiv \frac{2U(\sg_0)}{\sg_0^2}.
                 \end{equation}
                 If $\omega_0^2<0$ this is automaticaly satisfied. To understand this condition for $\omega_0^2 \geq 0$ note first that as the particle attains the origin its energy, $E_{part} \equiv \meio (\frac{d\sg}{dr})^2 - U_{\omega}(\sg)$, is equal to zero. From the equation of motion we know that the particle's energy is a decreasing function of the time $r$, in virtue of the action of the viscous force. The potential energy must therefore be positive at time $r=0$, when the particle starts at rest with position $\sg(0)\neq 0$. But if $\omega^2 < \omega_0^2$ the potential energy $-U_{\omega}$ is never positive. This is easy to see in fig.\ref{fig:1}, but can also be shown mathematically: If for some $\tilde{\sg}$ the potential is positive and $\omega^2 < \omega_0^2$ we would have
                 \bequ\begin{split}
                          & -U_{\omega}(\tilde{\sg})>0 \Rightarrow U(\tilde{\sg})<\meio\omega^2\tilde{\sg}\\
                          & \Rightarrow\frac{2U(\tilde{\sg})}{\tilde{\sg}^2}<\omega^2<\text{min}\left(\frac{2U(\sg)}{\sg^2}\right).
                 \end{split}\end{equation}
\end{itemize}

In this section we have seen that as long as $\omega_{0}^2 \neq m^2$, the theory contains classical solutions of the equations of motion, which are energy extrema in a sector of fixed charge. These solutions rotate in internal $U(1)$-space with frequency $\omega$ ($\omega_{0}^2 < \omega^2 < m^2$) and their modules $\sg(r)$ have a bounce-like shape, i.e. they have a maximum at $r=0$ and decrease monotonically towards zero at infinity.

\section{Stability of Q-balls}
\label{sec:3}

In the following paragraphs we will argue that the stability of Q-ball solutions depends on the way the charge changes with the internal frequency $\omega$. If at least some deviations from the solution grow with time, we say that the solution is unstable. If any perturbation remains oscillating around the solution, this is a stable one. It's clear that, due to charge conservation, minima of the energy for fixed charge are stable under small perturbations.

The question if a Q-ball solution is an {\bf absolute} minimum of the energy at fixed charge was answered by Coleman~\cite{col85}. He proved that as long as 
\bequ
                     E_{Qball} < m\,Q \text{, and } \omega_{0}^2>0
\end{equation}
the Q-ball is the absolute minimum configuration with charge $Q$.

There is however another more useful theorem regarding the {\bf local} stability of Q-balls, that also works for $\omega_{0}^2 \leq 0$, that is, in the case that \emph{the symmetry is broken}. It states that\footnote{This theorem was stated for a similar kind of non-topological solitons in ref.\cite{lee76}.}
\begin{itemize}
              \item if $\frac{\omega}{Q}\frac{dQ}{d\omega} < 0$, the Q-ball is a local minimum, and therefore stable;\\
              \item if $\frac{\omega}{Q}\frac{dQ}{d\omega} > 0$, the Q-ball is a saddle point of the energy, with one instability mode.   
\end{itemize}

To prove this we must investigate general perturbations of the solutions around the form $\phi = \bar{\sg}(r)\,e^{i\omega t}/\sqrt{2}$ which satisfy the eq.~of motion \eqref{um}. Since the general configurations $\sg({\bf x},t)\,e^{i\vartheta({\bf x},t)}/\sqrt{2}$ depend on both space and time it seems dificult to fix the charge in an \emph{explicit} way when performing variations of the energy. That is, we would like to find an expression, like eq.\eqref{EQ}, where the charge dependence is explicit, but which is valid for general configurations. We can solve this problem by introducing the following function,
\bequ\label{144}
               \omega(t) \equiv \dfrac{\intx \sg^2({\bf x},t)\,\dot{\vartheta}({\bf x},t)}{\intx \sg^2({\bf x},t)}=\frac{Q[\sg,\vartheta]}{I[\sg]},
\end{equation}
and spliting $\vartheta({\bf x},t)$ as
\bequ
                \vartheta({\bf x},t)=\theta_0(t) + \theta({\bf x},t), 
\end{equation}
where $\theta_0(t)$ is an indefinite integral of $\omega(t)$, which depends only on time\footnote{$\theta({\bf x},t)$ is constrained by $\intx\sg^2\dot{\theta}=0$ (see eq.\eqref{144}), but this is irrelevant for the discussion which follows.}. With these definitions the energy functional reads:
\bequ\begin{split}
                 E[\sg,\vartheta] & = \intx \left[\frac{1}{2}(\dot{\sg})^2 + \frac{1}{2}\sg^2(\dot{\theta})^2  + \meio\sg^2(\nabla\theta)^2 \right]\\                                                          & \quad + \frac{1}{2}\frac{Q^2}{I[\sg]} + \int d^3x\left[\frac{1}{2}(\nabla\sigma)^2 + U(\sigma) \right]\\                                                                  & \equiv K[\sg,\theta] + E_Q[\sg],
\end{split}\end{equation}
where $E_Q$ is the functional defined by eq.\eqref{EQ}. For a configuration of the Q-ball type, $\vartheta({\bf x},t)=\omega t$ and $\sg({\bf x},t)=\bar{\sg}(r)$. We have thus $\theta({\bf x},t)=const.$ and therefore it follows readily that $K_{Qball} = 0$. Since in general $K[\sg,\theta] \geq 0$, Q-balls (or configurations of the Q-ball type) are minima of this functional. This means that the stability depends only on the behaviour of the functional $E_Q[\sg]$ under variations $\sg({\bf x},t) = \bar{\sg}(r) + \delta\sg({\bf x},t)$. Note that for this variational problem only the spatial dependence of $\sg$ is relevant, the time $t$ playing the role of a fixed parameter. This proves the assertion of Section 2 that \emph{the stability of Q-balls depends only on the behaviour of the energy in the subspace of configurations of the type $\sg({\bf x})\,e^{i\omega t}$}, for which the energy reduces to $E_Q[\sg]$. 

The remaining task is, therefore, to prove that the above theorem is true in this restricted subspace. To do this we will first show that \emph{in addition to the 3 translational zero-modes $\partial_i \bar{\sg}$ of $E_Q[\bar{\sg}]$ there is only one other zero-mode if and only if $\frac{\omega}{Q}\frac{dQ}{d\omega}=0$}: From eq.\eqref{cinco} we see that any null mode $\psi$ must satisfy
\bequ\label{29}
                    \mathcal{H}\psi\equiv\meio h\psi + \frac{2\omega^2}{I}\bar{\sg} \intx \bar{\sg}\psi=0,
\end{equation}
where $h \equiv -\nabla^2 + U_{\omega}''(\bar{\sg})$. With $A \equiv -\frac{4\omega^2}{I} \intx \bar{\sg}\psi$ we get
\bequ
                       h\psi = A\bar{\sg}.
\end{equation}
Differentiating the equation of motion with respect to $\omega$ one sees that the last equation is satisfied by $\psi=\frac{d\bar{\sg}}{d\omega}$ with $A=2\omega$. For any other solution $\psi$ we have thus
\bequ
                       h\left(\psi - \frac{A}{2\omega}\frac{d\bar{\sg}}{d\omega}\right)=0,
\end{equation}
and since the $\partial_i \bar{\sg}$ are the only functions for which $hf=0$ we see that $\psi$ is allways a linear combination of the $\partial_i \bar{\sg}$ and of $\frac{d\bar{\sg}}{d\omega}$. Now, if $\frac{d\bar{\sg}}{d\omega}$ is a solution of eq.\eqref{29} we get
\bequ
                       1+\frac{2\omega}{I}\intx\bar{\sg}\frac{d\bar{\sg}}{d\omega}=0,
\end{equation}
and therefore
\bequ
                       \frac{\omega}{Q}\frac{dQ}{d\omega}=\frac{\omega}{Q}\left[\frac{Q}{\omega}+2\omega\intx\bar{\sg}\frac{d\bar{\sg}}{d\omega}\right]=0,
\end{equation}
as we intended to prove.

We will now show that the appearence of the \emph{extra} null mode when $\frac{\omega}{Q}\frac{dQ}{d\omega}=0$ really signals the change of sign of an eigenvalue of $\mathcal{H}$ and with it the transition between stability and instability. The first step is to prove that since $h$ has only one negative eigenvalue \emph{$\mathcal{H}$ can have at most one negative eigenvalue}: If $\phi_i$ is a negative mode of $\mathcal{H}$ it must satisfy 
\bequ
                \intx \phi_i \delta\sg_{-1} \neq 0,
\end{equation}
where $\delta\sg_{-1}$ is the negative mode of $h$, or else $\int \phi_i \mathcal{H} \phi_i \equiv \lambda_i$ would be positive. If there are more than one negative mode of $\mathcal{H}$ we can build a suitable linear combination $\Phi=\sum a_i\phi_i$ for which $\intx \Phi \, \delta\sg_{-1} = 0$. But this would mean that
\bequ\begin{split}
                 0 & >\sum a_i^2 \lambda_i = \intx \Phi \mathcal{H} \Phi ={}\\
                                        & \quad = \meio\intx \Phi h\Phi + \frac{2\omega^2}{I}\left(\intx \bar{\sg}\Phi\right)^2>0,
\end{split}\end{equation}
what is impossible. There can be therefore at most one negative eigenvalue.

It remains to show that if $\lambda(\omega)$ is the \emph{extra} eigenvalue of $\mathcal{H}$ which is zero when $\frac{dQ}{d\omega}=0$, we have \emph{$\lambda(\omega)>0$ when $\frac{dQ}{d\omega}<0$ and vice-versa}. If $\psi(\omega)$ is the eigenvector corresponding to $\lambda(\omega)$, and $\bar{\omega}$ is defined by $\lambda(\bar{\omega})=0$, we have (up to a multiplicative constant) $\psi(\bar{\omega})=\frac{d\bar{\sg}}{d\bar{\omega}}$. Differentiating $\mathcal{H}\psi=\lambda\psi$ at $\omega=\bar{\omega}$ we get thus
\bequ
                \frac{d\mathcal{H}}{d\bar{\omega}}\frac{d\bar{\sg}}{d\bar{\omega}}+\mathcal{H(\bar{\omega})}\frac{d\psi}{d\bar{\omega}}=\frac{d\lambda}{d\bar{\omega}}\frac{d\bar{\sg}}{d\bar{\omega}}.
\end{equation}
We multiply this equation with $\frac{d\bar{\sg}}{d\bar{\omega}}$ and integrate to obtain
\bequ
                 \intx\frac{d\bar{\sg}}{d\bar{\omega}}\frac{d\mathcal{H}}{d\bar{\omega}}\frac{d\bar{\sg}}{d\bar{\omega}}=\frac{d\lambda}{d\bar{\omega}}\cdot\intx\left(\frac{d\bar{\sg}}{d\bar{\omega}}\right)^2.
\end{equation}
However we can rewrite the l.h.s. of this equation as 
\bequ
                 \frac{d}{d\bar{\omega}}\intx\frac{d\bar{\sg}}{d\omega}\mathcal{H}\frac{d\bar{\sg}}{d\omega} -2\intx\frac{d^2\bar{\sg}}{d\bar{\omega}^2}\mathcal{H}(\bar{\omega})\frac{d\bar{\sg}}{d\bar{\omega}}\equiv\frac{dF}{d\bar{\omega}},
\end{equation}
where $F(\omega)=\intx\frac{d\bar{\sg}}{d\omega}\mathcal{H}\frac{d\bar{\sg}}{d\omega}$ is given by (see eq.\eqref{cinco})
\bequ\begin{split}
                 F(\omega) & =\omega\intx\bar{\sg}\frac{d\bar{\sg}}{d\omega}\left[1+\frac{2\omega}{I}\intx\bar{\sg}\frac{d\bar{\sg}}{d\omega} \right]\\
                                                               & =\meio\left[\frac{dQ}{d\omega}-\frac{Q}{\omega}\right]\frac{\omega}{Q}\frac{dQ}{d\omega}.
\end{split}\end{equation}
Since $\frac{dQ}{d\omega}(\bar{\omega})=0$ we see that in a small neighbourhood of $\bar{\omega}$ we have $\text{\emph{sign}}(F(\omega))=\text{\emph{sign}}(\lambda(\omega))=-\text{\emph{sign}}(\frac{dQ}{d\omega})$. But this is enough to prove that $\text{\emph{sign}}(\lambda(\omega))=-\text{\emph{sign}}(\frac{dQ}{d\omega})$ for any values of $\omega$.

%It remains only to show that $\frac{\omega}{Q}\frac{dQ}{d\omega}>0$
%is related to instability (and therefore $\frac{\omega}{Q}\frac{dQ}{d\omega}<0$
%with stability). This is simple: put $\delta\sg=\frac{d\bar{\sg}}{d\omega}$ in eq.\eqref{cinco}. We get
%\bequ\begin{split}
%                       \delta^2E_Q[\tfrac{d\bar{\sg}}{d\omega}] & =\omega\intx\bar{\sg}\frac{d\bar{\sg}}{d\omega}\left[1+\frac{2\omega}{I}\intx\bar{\sg}\frac{d\bar{\sg}}{d\omega} \right]\\
%                                                               & =\meio\left[\frac{dQ}{d\omega}-\frac{Q}{\omega}\right]\frac{\omega}{Q}\frac{dQ}{d\omega}.
%\end{split}\end{equation}
%If we choose $\frac{dQ}{d\omega}>0$ but small enough we get $\delta^2E_Q<0$. This must remain true untill the negative eigenvalue becomes positive, that is, untill $\frac{\omega}{Q}\frac{dQ}{d\omega}<0$.

%We would like to emphasize that it seems that some of the results of this and the last sections are widely unknown or forgotten in the literatur, with the result that some \emph{wrong} considerations were produced and even some numerical results were \emph{adapted} to wrong assumptions~\cite{axe00}. For instance, compare eq.\eqref{116} with eq.(8) of reference~\cite{kus97} and the following comments. The author is not aware of the existence of contributions to the energy's second variation which are non-local in the fluctuations.

\section{Thin-wall and thick-wall regime}
\label{sec:4}

As we have seen, if we want to know whether a Q-ball solution is stable
or not, instead of solving two second order partial differencial
equations in $3+1$ dimensons, we need only to know whether
$\frac{dQ}{d\omega}$ is positive or negative. There are two limiting
cases where we can apply these results without making use of a computer:
when $\omega^2 \to \omega^2_0 \geq 0$, and when $\omega^2 \to m^2$. The
first limit is known as the thin-wall regime while the second one as the
thick-wall regime.
\subsection{Thin-wall approximation.}
\label{sec:4.1}

In the mechanical analogon which was described in sec.\ref{sec:2} the
initial position must be chosen in such a way that for $r \to \infty$ it
won't undershoot or overshoot the top of the hill at $\sg = 0$.

When $\omega^2_0 \geq 0$~\footnote{~$\omega^2_0 \geq 0$ is the condition that the classical potencial $U(\sg)$ is not negative. The case with $\omega^2_0 < 0$ will be analyzed in sec.4.3.} and $\omega^2 \to \omega^2_0$, the absolute ma\-xi\-mum of $-U_{\omega}$ becomes virtually degenerate with the one at $\sg = 0$. In this limit the particle must spend a very long time close to $\sg_0$ (defined by eq.\eqref{omegazero}) otherwise it would undershoot the origin $\sg = 0$. To see this note first that if $\sg_1$ is the zero of $U_{\omega}$ and $\sg(0) < \sg_{1}$, the particle doesn't have enough energy to reach the origin, for the energy is a decreasing function of time. Now, if $\sg(0) > \sg_{1}$, $\sg(0)$ is so close to $\sg_{0}$ that we can linearize the eq. of motion:
\bequ
                    \left [ \frac{d^2}{dr^2} + \frac{2}{r}\frac{d}{dr} - \mu^2 \right ](\sg(r) - \sg_{0}) = 0,
\end{equation}
with $\mu^2 \equiv U_{\omega}^{''}(\sg_{0}) > 0$. This equation is a good approximation as long as $s(r)\equiv\sg_{0}-\sg(r)$ is not too large. To find the solution with $\frac{d\sg}{dr}(0)=0$ is quite a simple task:
\bequ\label{solu}
                     s(r) = s_0\frac{\sinh{(\mu r)}}{\mu r}.
\end{equation}
We still have the freedom of choosing $\sg(0)$ but since we know that $\sg(0) > \sg_{1}$ we see that as $\omega^2 \to \omega_{0}^2$, $s_0 = (\sg_{0} - \sg(0)) \to 0$. When this happens eq.\eqref{solu} implies that the particle spends more and more time close to $\sg_{0}$. We can define the \emph{time} R that the particle spends close to $\sg_{0}$ to satisfy
\bequ
                     s(R) = \sg_0-\sg(R) = s_R,
\end{equation}
where $s_R$ is \emph{just} small enough to allow for the linear approximation. Now, since $\sinh(x)/x$ is a growing function of $x$ it's clear that as $s_0 \to 0$ (i.e. $\omega^2 \to \omega_{0}^2$) we must expect $R\mu \to \infty$. On the other hand $\mu^2 =-\omega^2 + const.$ is bounded in $[\omega_0^2, m^2]$, which proves that in the thin-wall limit, as $s_0=\sg_{0}-\sg(0)$ gets smaller, $R$ becomes larger and larger and grows to infinity. Note that the asymptotic behaviour of $R$ is independent of the definition point $s_R$.

We can also easily show that for $r=R$ the damping term is already unimportant when compared with the potential term\footnote{We use here and in eq.\eqref{eq:45} $\sinh(\mu R)\simeq\cosh(\mu R)\simeq\meio e^{\mu R}$.}:
\bequ\label{162}
                       \left|\frac{1}{R}\frac{d\sg}{dr}(R)\right|  \simeq \frac{s_0}{R}\mu\frac{e^{\mu R}}{2\mu R} \simeq \frac{1}{\mu R}\mu^2 s_R \ll \left|U_{\omega}^{'}(\sg(R))\right|.
\end{equation}
That means that we can describe the Q-ball for $\omega^2 \to \omega_{0}^2$ as
\bequ
                       \sg(r) = \left\{ \begin{array}{ll}
                                            \sg_0 - s(r) & \text{if $r<R$}\\
                                            \bar{\sg}(r-R) & \text{if $r>R$}
                                        \end{array} \right.
\end{equation}
where $\bar{\sg}(x)$ is the solution of the eq. of motion \emph{without damping term}, which fits to $\bar{\sg}(0)=\sg_0 - s_R$ and $\bar{\sg}(\infty)=0$. Back to the field-theoretical language we can describe this Q-ball as formed by a very large \emph{core} with radius $R$ surrounded by a comparably thin surface. The free parameter $\sg(0)$ (or $s_0$ or $R$) must now be adjusted as to minimize the energy at fixed $Q$ or $S_{\omega}$ at fixed $\omega$. This is not a difficult thing to do.

We must first note the following: (i) for $r<R$ we have $U_{\omega}(\sg)=\meio\mu^2s^2-\epsilon$, where $\epsilon=-U_{\omega}(\sg_0)=\meio\sg_0^2(\omega^2-\omega_0^2)$ is the energy difference between the tops of the two \emph{hills}; (ii) when $r>R$ the eq. of motion has no damping term and therefore $\meio(\frac{d\bar{\sg}}{dr})^2=U_{\omega}(\bar{\sg})$ (i.e. the \emph{particle's energy} is conserved). A straightforward calculation gives
\bequ
                         S_{\omega}=S_{\omega}^{<}+S_{\omega}^{>},
\end{equation}
where the core's contribution is
\bequ\label{eq:45}
                         S_{\omega}^{<}=-\frac{4\pi}{3}R^3\epsilon + 2\pi R^2 s_R^2 \mu - 2\pi Rs_R^2,
\end{equation}
and
\bequ
                         S_{\omega}^{>}=4 \pi R^2 T +2RA_1+A_2,\quad T\equiv\int_{0}^{\sg_R}d\sg\sqrt{2U_{\omega}(\sg)}
\end{equation}
is the surface's contribution. The quantities $T$ and $A_n \equiv 8\pi\int_{0}^{\infty}dx x^n U_{\omega}(\bar{\sg})$ depend weakly on $\omega$ when compared with any positive power of $R$, which as we know become infinite as $\omega^2 \to \omega_0^2$. In this limit we can drop powers of $R$ of order lower than $2$, obtaining therefore
\bequ\label{eq:47}
                         S_{\omega}=-\frac{4\pi}{3}R^3\epsilon + 2\pi(2T+s_R^2\mu)R^2.
\end{equation}
The condition that we can neglect the term linear in $R$ is $R \gg A_1/2\pi T$. A quick look at the definitions of $T$ and $A_1$ shows that $\delta \equiv A_1/2\pi T$ is the \emph{thickness} of the Q-ball's wall. This is the reason for calling this the thin-wall limit.

The quantity $\meio\tau\equiv T+\meio s_R^2\mu$ has also a simple interpretation: If we note that $s_R^2\mu= 2 \int_{0}^{s_R} ds \sqrt{s^2\mu^2}=2 \int_{\sg_R}^{\sg_0} d\sg \sqrt{2U_{\omega_0}}$ it becomes clear that
\bequ
                         \meio\tau = \int_{0}^{\sg_R} d\sg \sqrt{2U_{\omega}}+\int_{\sg_R}^{\sg_0} d\sg \sqrt{2U_{\omega_0}} \simeq \int_{0}^{\sg_0} d\sg \sqrt{2U_{\omega_0}}
\end{equation}
is the \emph{surface tension} of the thin-walled Q-ball.

The function of eq.\eqref{eq:47} has a maximum at $\bar{R}=\tau/\epsilon$. Putting this back in \eqref{eq:47} and using eq.\eqref{eq:14},\eqref{193} we get
\bequ
                         S(\omega)=\frac{2\pi}{3}\frac{\tau^3}{\epsilon^2}=\frac{8\pi}{3}\frac{\tau^3}{\sg_0^4}\frac{1}{(\omega^2-\omega_0^2)^2},
\end{equation}
\bequ\label{eq:50}
                         Q(\omega) =-\frac{dS}{d\omega} = \frac{32\pi}{3}\frac{\tau^3}{\sg_0^4}\frac{\omega}{(\omega^2-\omega_0^2)^3}=\frac{4\pi}{3}\bar{R}^3\sg_0^2\omega,
\end{equation}
and
\bequ
                         E(\omega)=S(\omega)+\omega Q(\omega)=\omega Q(\omega)\left[\frac{5}{4}-\frac{\omega_0^2}{4\omega^2}\right].
\end{equation}
Assuming a weak $\omega$-dependence for $\tau$, it follows from eq.\eqref{eq:50} the stability of thin-walled Q-balls:
\bequ
                         \frac{\omega}{Q}\frac{dQ}{d\omega}=1-6\frac{\omega^2}{\omega^2-\omega_0^2}<0.
\end{equation}
These equations are valid for both $\omega_0^2>0$ and $\omega_0^2=0$ and show a difference in the properties of thin-walled Q-balls between models with \emph{unbroken symmetry} ($\omega_0^2>0$) and \emph{broken symmetry} ($\omega_0^2=0$):
%\footnote{Part of these results were obtained in \cite{col85,spec87} with a different approach. The advantage of our method is: (i) we \emph{can} identify the point where the approximation breaks; (ii) we have a better knowledge of $Q(\omega)$ in the $\omega_0^2>0$ case; (iii) our formula can give \emph{quantitative} results for concrete models. We just need to calculate $\tau$ etc.}
\begin{itemize}
                    \item [(i)] For $\omega_0^2>0$ and $\omega^2\simeq\omega_0^2$ we get the Q-balls as they were first conceived by Coleman~\cite{col85}. They have the property that $E=\omega_0 Q$, i.e. they behave like a ball of \emph{Q-matter}: the energy being proportional to the charge which is proportional to the volume (of the core).\\

                    \item [(ii)] For $\omega_0^2=0$ and in the thin-wall limit, we get from eq.\eqref{eq:50} $Q\sim \omega^{-5}$ and
                    \bequ
                                    E=\frac{5}{4}\left(\frac{32\pi}{3}\frac{\tau^3}{\sg_0^4}\right)^{1/5} Q^{4/5}.
                    \end{equation}
                                 This shows that such Q-balls can't be seen as Q-matter \cite{spec87}. The charge doesn't grow with the volume but as $Q\sim \bar{R}^{5/2}$ and the energy as $E\sim\bar{R}^2$. The reason for this behaviour is simple: The core of the Q-ball is nearly in the asymetric vacuum and it's energy density decreases exponentially with its radius $\bar{R}$. The energy of the core becomes thus less important than the surface's tension.\\

                    \item [(iii)] Finally, if $\omega_0^2>0$ but $m^2 \gg \omega_0^2$ the approximations we made are still valid if $m^2 \gg \omega^2 \gg \omega_0^2$. In this regime, although we have $\omega_0^2 \neq 0$, we can use the results of (ii) since $\omega_0^2/\omega^2 \simeq 0$. The results of (i) naturally still apply as $\omega^2 \to \omega_0^2$.
\end{itemize}

\subsection{Thick-wall approximation.}
\label{sec:4.2}

We want now to look at theories which can be put in the form
\bequ\label{64}
                       U(\sg)=\frac{1}{2}m^2\sg^2-A\sg^n + \sum_{p>0} B_p\sg^{n+p},
\end{equation}
with $A$ a positive quantity. In the limit $\omega^2 \to m^2$, the Q-ball cannot in general be approximated by a step function as before. This limit is called, therefore, the thick-wall limit. Some authors (see ref.\cite{kus97,lee91,mul00}) proposed that since $\sg(0) \to 0$ we may neglect the terms with powers higher than $n$ in eq.\eqref{64}, when calculating the properties of the Q-ball. In this way the energy and the charge of the Q-balls get a simple dependence on the relevant parameters of the theory, and the Q-balls stability can be analyzed. This is what we are going to investigate in the following lines.

Define
\bequ\label{eq:56}
                        \varepsilon_{\omega} \equiv m^2 - \omega^2.
\end{equation}
and rescale the field as
\bequ\label{sglinha}
                    \sg' \equiv \frac{\sg}{(\varepsilon/A)^{\frac{1}{n-2}}}.
\end{equation}
Changing the variable $r$ as $r \to r\varepsilon^{\frac{1}{2}}$ we obtain (eq.\eqref{115})
\bequ
                  S_{\omega}[\sg] = A^{-\frac{2}{n-2}}\varepsilon^{ \frac{2}{n-2}-\frac{1}{2}}S_{\varepsilon}'[\sg'],
\end{equation}
where
\bequ\begin{split}\label{eq:56a}
                  S_{\varepsilon}'[\sg'] & \equiv \int d^3r\,\left[ \frac{1}{2}\left(\frac{d\sg'}{dr}\right)^2 + \frac{1}{2}(\sg')^2 - (\sg')^n \right]\\
                           & \quad + \int d^3r\,\sum_p \frac{B_p}{A}\left(\frac{\varepsilon}{A}\right)^{\frac{p}{n-2}}(\sg')^{n+p}.
\end{split}\end{equation}

Suppose we know that the Q-ball solution $\bar{\sg}$ is an extremum of $S_{\omega}[\sg]$. To this solution corresponds an extremum of $S_{\varepsilon}'[\sg']$, which is obtained by using eq.\eqref{sglinha}. We now hope that $\bar{\sg}'(0)$ as a function of $\varepsilon_{\omega}$ falls quickly enough as $\varepsilon_{\omega} \to 0$ so that the terms of order higher than $n$ become irrelevant to the calculation of $\bar{\sg}'$. In that case $S_{\varepsilon}'[\bar{\sg}'] \equiv S_n$ doesn't depend on $\varepsilon$, and we get the following simple expression for $S(\omega)\equiv S_{\omega}[\bar{\sg}]$
\bequ
                  S(\omega) = A^{-\frac{2}{n-2}}(m^2-\omega^2)^{ \frac{2}{n-2}-\frac{1}{2}}S_n.
\end{equation}
We can now calculate the charge and the energy from this expression. We have thus (for $n>2$):
\bequ\begin{split}\label{195}
               Q(\omega) & =-\frac{dS(\omega)}{d\omega}={}\\
                         & =  A^{-\frac{2}{n-2}}S_n\left(\frac{2}{n-2}-\frac{1}{2}\right)2\omega(m^2-\omega^2)^{ \frac{2}{n-2}-\frac{3}{2}},
\end{split}\end{equation}
and
\bequ\begin{split}
               E(\omega) & =A^{-\frac{2}{n-2}}S_n(m^2-\omega^2)^{ \frac{2}{n-2}-\frac{3}{2}}\\
                         & \quad \quad \cdot\left[(m^2-\omega^2) + 2\omega^2\left(\frac{2}{n-2}-\frac{1}{2}\right)\right].
\end{split}\end{equation}\\
Something must be wrong for $n\geq 6$: If $n>6$ and $\omega>0$, one sees that $Q, E \to - \infty$. That's obviously impossible as both the energy and the charge should be positive. Also for $n=6$ these equations give $E(\omega)=A^{-\frac{1}{2}}S_6\neq 0$ and $Q(\omega)=0$, but this last expression implies that $\sg(r)=0$ with the consequence that $E(\omega)=0$!

We must conclude that, at least for $n \geq 6$, this so-called thick-wall approximation is a bad approximation. The reason is simple: The approximation we made leads us to a functional with no extrema for $n \geq 6$, so that, in this case, $S_n$ does not exist. To prove this we start with the (false) assumption that, for $n \geq 6$ the functional
\bequ
                       S[\sg({\bf x})] = \intx \left(\meio(\nabla\sg)^2 + \meio \sg^2 - \sg^n\right),
\end{equation}
has an extremum $\bar{\sg}({\bf x})$. As a consequence, the functions $X(\alpha) \equiv S[\sg({\bf x}/\alpha)]$ and $Y(\lambda) \equiv S[\lambda\sg({\bf x})]$ should have an extremum at $\alpha = 1$ and $\lambda = 1$ respectively. But this means that
\bequ
                    X'(1) = \intx \meio(\nabla\sg)^2 + 3 \intx \left(\meio \bar{\sg}^2 - \bar{\sg}^n\right) = 0,
\end{equation}
and
\bequ
                     Y'(1) = \intx \left(\meio(\nabla\sg)^2 + \meio \bar{\sg}^2\right)-\frac{n}{2}\intx \bar{\sg}^n = 0.
\end{equation}
Subtracting the first expression from the second one we thus have
\bequ
                       \intx \bar{\sg}^2 = \frac{6-n}{2}\intx \bar{\sg}^n.
\end{equation}
For $n \geq 6$ and $\bar{\sg}>0$ this is obviously wrong, as we intended to show.

Thus we have proved that the so-called thick-wall approximation is useful only for $n<6$. From eq.\eqref{195} we get
\bequ\label{eq:64}
                        \frac{\omega}{Q}\cdot\frac{dQ}{d\omega} = 1 -2\left(\frac{2}{n-2} - \frac{3}{2}\right)\frac{\omega^2}{m^2 -\omega^2},
\end{equation}
so that as $\omega^2 \to m^2$ we have thus
\begin{itemize}
         \item \quad $\dfrac{\omega}{Q}\cdot\dfrac{dQ}{d\omega}>0$ for $n=4,5$,
         \item \quad $\dfrac{\omega}{Q}\cdot\dfrac{dQ}{d\omega}<0$ for $n=3$.
\end{itemize}
This means that thick-walled Q-balls are known to be stable only if $n=3$.

\subsection{An application: Q-balls in the false vacuum.}
\label{sec:4.3}

As an illustration of the use of the approximations we have just discussed, we will investigate what happens when the symmetric vacuum becomes unstable, i.e. for $\omega_0^2<0$. Consider the following potential
\bequ
                U(\sg)=\meio m^2\sg^2 -\frac{\alpha}{3}\sg^3+\frac{\lambda}{4}\sg^4.
\end{equation}
We explained in sec.\ref{sec:2} that the energy spectrum includes Q-balls if $\omega_0^2=m^2-\frac{2}{9}\frac{\alpha^2}{\lambda}<m^2$ (see eq.\eqref{omegazero}), i.e. if $\lambda>0$. For $\omega_0^2 \geq 0$, the results of \ref{sec:4.1} and \ref{sec:4.2} show that both thick-walled and thin-walled Q-balls are stable, which is enough to show, using Coleman's theorem, that \emph{all} Q-balls are absolutely stable configurations in the unbroken phase. 

The transition $\omega_0^2 \geq 0 \to \omega_0^2<0$ changes the energy spectrum in a dramatic way\cite{spec87,kus97a}: If before we had stable Q-balls with all possible charges and energies, there is now a maximum charge $Q_c$ and a maximum energy $E_c$ which Q-balls can have. One way of seeing this is using the thin-wall approximation of sec.\ref{sec:4.1}. As we will show below, we still can use this approximation when $-\omega_0^2 \ll m^2$ and $\omega^2 \ll m^2$. Eqs.\eqref{eq:50} ff show that there is a frequency $\omega_c^2\equiv \frac{1}{5}|\omega_0^2|$ such that $Q(\omega)$ attains a maximum at $\omega=\omega_c$ and only Q-balls with $\omega^2 >\omega_c^2$ are stable. Within this approximation we get
\bequ
                     Q_c=\frac{32\pi}{3\sqrt{5}}\left(\frac{5}{18}\right)^3\sg_0^2\frac{\sqrt{m^2+|\omega_0|^2}^3}{|\omega_0|^5},
\end{equation}
\bequ
                     E_c=\frac{\sqrt{5}}{2}|\omega_0|Q_c,
\end{equation}
and
\bequ
                     {\bar R}_c=\frac{5}{9}\frac{\sqrt{m^2+|\omega_0|^2}}{|\omega_0|^2},
\end{equation}
where we used $\tau=\frac{1}{3}\sqrt{m^2+|\omega_0|^2}\sg_0^2$. As we know, this approximation is valid as long as the radius of the \emph{critical} Q-ball, ${\bar R}_c$, is much larger than the wall's thickness, $\delta\sim \sqrt{m^2-\omega_0^2}^{-1}$, that is for $m^2 \gg |\omega_0|^2$, as we said above.

The question is now: \emph{What happens as $|\omega_0|^2$ becomes larger, that is, for a deeper true vacuum}? We will show that for $|\omega_0|^2 > 4m^2$ the critical Q-ball is already a thick-walled Q-ball. For the thick-wall approximation to be valid, the term $\sim (\sg')^{n+1}=(\sg')^4$ in eq.\eqref{eq:56a} should be neglegible when compared to the term $\sim (\sg')^n=(\sg')^3$. Using the numerical value $\sg'(0)^2 \simeq 2$ we get the following condition on the parameters: $m^2+|\omega_0|^2\gtrsim 10(m^2-\omega^2)$. This shows that if $|\omega_0|^2 > 4m^2$, the approximation is valid for $\omega^2\geq \meio m^2$. Now, a short look at eq.\eqref{eq:64} shows that if the critical Q-ball is thick-walled, the critical frequency is precisely $\omega_c^2=\meio m^2$. Furthermore we have
\bequ
                     Q_c=\frac{3}{2}S_3 m^2 \frac{\sg_0^2}{(m^2+|\omega_0|^2)^2},
\end{equation}
\bequ
                     E_c=\frac{2\sqrt{2}}{3}mQ_c.
\end{equation}
In conclusion, both thin-wall and thick-wall approximations show, for $|\omega_0|^2<m^2/100$ and $|\omega_0|^2>4m^2$, resp., that there is a maximum charge $Q_c$ which stable Q-balls can have. Although we can't calculate $Q_c$ and $\omega_c^2$ analiticaly in the range $|\omega_0|^2 \in [m^2/100, 4m^2]$, using \emph{scaling} properties of the model it can be shown that the picture remains the same\cite{pac00}. 

%We can namely show that 
%\bequ
%                   Q_c=\frac{\sg_0^2}{\sqrt{m^2+|\omega_0|^2}}f(|\omega_0|^2),
%\end{equation}
%where $f(|\omega_0|^2)$ satisfies the equation
%\bequ
%                   \frac{df}{d|\omega_0|^2}=-\frac{f}{2\omega_c^2}.
%\end{equation}
%This proves that $Q_c$ always decreases as the asymmetric vacuum becomes deeper.

\section{Quantum corrections}
\label{sec:5}

This section is meant to explain why Q-balls, which are \emph{classical} configurations, are important to quantum theory. As we will see, in certain circumstances the energy of a Q-ball of charge $q$ is the \emph{zeroth-order} contribution in a semi-classical expansion to the energy of the lowest lying state of charge $q$. To show this it is usefull to investigate the following \emph{partition function}\footnote{A few words on the notation we use: $|\phi\rangle=|\phi({\bf x})\rangle$ is the eigenvector of the field operator $\Phi({\bf x})$ with eigenvalue $\phi({\bf x})\equiv\phi_R({\bf x})+i\phi_I({\bf x})$. The measures $d\phi$ and $[d\phi]$ are respectively $\sim \prod_{{\bf x}} d\phi_R({\bf x})d\phi_I({\bf x})$ and $\sim\prod_{{\bf x},t} d\phi_R({\bf x},t)d\phi_I({\bf x},t)$, resp.}:
\bequ
                     Z(T) = \text{tr} [e^{-HT}] = \int d\phi\,\langle \phi|\,  e^{-HT}\, |\phi\rangle,
\end{equation}
for U(1)-invariant scalar theories. It is possible to perform a separation of the contributions of sectors of different charge to $Z(T)$
\bequ\label{318}
                       Z(T) = \sum_{q}\, \int d\phi\, \int_{0}^{2\pi} \frac{d\alpha}{2\pi}\, e^{-i\alpha q}\, \langle \phi|\, e^{i\alpha Q}\, e^{-HT}\, |\phi\rangle.
\end{equation}
This can be justified in a \emph{heuristic} way~\cite{raj1,raj}, by noting that
\bequ
                       \int_{0}^{2\pi} \frac{d\alpha}{2\pi}\, e^{i\alpha (Q-q)} = \delta_{Q,q}.
\end{equation}
(For a more formal derivation see ref.\cite{ben89}.) In the limit $T \to +\infty$, we have
\bequ
                       \lim_{T \to \infty} Z(T) = \sum_{q}\,e^{-E_q^0 T} \equiv \sum_{q}\, Z_0^q (T) ,
\end{equation}
where $E_q^0$ is the lowest energy eigenvalue in the sector of charge $q$. As is shown in the appendix, $Z_0^q (T)$ turns out to be given by
\bequ\begin{split}\label{funcint}
                       Z_0^q (T) = \int\,&[d\phi]\,\delta\left((q-Q)/\sqrt{I}\right)\exp{\left(-\int dt\,E \right)}\\
                                & \cdot\cos{\left(q\int\,dt\frac{Q}{I}-q\alpha\right)},
\end{split}\end{equation}
where the integration is made over configurations whose values at $t=T/2$ and $t=-T/2$ differ only by a global phase, i.e $\phi(T/2)=\phi(-T/2)e^{i\alpha}$ and $Q[\phi]$ and $I[|\phi|]$ are the functionals defined in sec.\ref{sec:2}.

The important feature in this expression is that the $\delta$-func\-tion constrains the integration to be over configurations of char\-ge $q$. As we know, there is, in certain cases, a configuration which minimizes $E[\phi]$ for the given charge: \emph{the Q-ball}. Furthermore, this configuration also makes the argument of the $cosinus$ zero, because $\alpha=\omega T=\int QI^{-1}dt$. To conclude: the (stable) Q-ball configuration makes the integrand of the above integral a \emph{maximum}. But this is not enough an argument for performing a semi-classical expansion around the Q-ball. The point is that if there are configurations with nearly the same energy, and with the same charge, as the Q-ball for which $q(\int dt\, Q/I-\alpha)=2\pi k,\,k \in \mathbb{Z}\backslash\{0\}$, the oscillating part of the integrand erases the contribution of the neighbourhood of the Q-ball. This can be the case, for instance, for thin-walled Q-balls, as these have a low-lying mode, with eigenvalue $(6\omega^2/(\omega^2-\omega_0^2)-1)R^{-1}$, which becomes a zero-mode as $R \to \infty$.

Only in the case the cosinus is not oscillating that fast in the neighbourhood of the Q-ball we can perform the semi-classical expansion around this configuration, in which case the classical energy is just the zero-order contribution to $E_q^0=T^{-1}\ln{Z_0^q(T)}$. We can then set the cosinus equal to 1, rewrite the $\delta$-function as \cite{cal75}
\bequ
                    \prod_{t}\delta((q-Q)/\sqrt{I})\sim\lim_{a \to 0} \exp{\left(-\frac{1}{2a}\int dt \frac{(q-Q)^2}{I}\right)},
\end{equation}
and then expand the energy functional around the Q-ball
\bequ
                    E[\phi]=E_{Qball}+E_{fl}[\psi],
\end{equation}
where $E_{fl}[\psi]$ is the energy contained in the fluctuations $\psi=\phi-\bar{\sg}\,e^{i\omega t}/\sqrt{2}$. In this way we get finally
\bequ\label{eq:76}
                    e^{-E_q^0 T}=e^{-E_{Qb}T}\lim_{a \to 0}\int [d\psi]\exp{\left[-\int dt \left(E_{fl}+\frac{Q_{fl}^2}{2aI_{fl}}\right)\right]},
\end{equation}
where $Q_{fl}[\psi]\equiv Q[\phi]-q$ and $I_{fl}[\psi]\equiv I[|\sg e^{i\omega t} + \sqrt{2}\psi|]$. Starting from this expression we can calculate the quantum corrections to the energy of the Q-ball by performing common perturbation theory.

\section{Conclusions}
\label{sec:7} 

In the body of this paper we presented an analytical
investigation of the main properties of Q-balls. We want now to
underline part of the results, some of them because they are new,
others as they seem to be unknown in the recent literature thus leading
to some incorrect statements and others because of their relevance.

As we have seen in sec.\ref{sec:2}, the existence of Q-balls, solutions
of non-zero charge, is a general property of scalar models with an
U(1)-symmetry: They exist in the case that $\omega_0^2=$
min$(2U(\sg)/\sg^2) < m^2$. As we also said, there is a theorem by
Coleman which states the \emph{absolute} stability of those Q-balls for
which $\omega_0^2>0$ and $E_{Qb}<mQ$. This is however of restricted use.
For instance, the gaussian-shaped B-balls which arise in mo\-dels with
gra\-vi\-ty-me\-dia\-ted SUSY-brea\-king have $\omega_0^2=0$ (the potential grows
too slow at the infinity). For this case and the one with $\omega_0^2<0$
we can use the theorem of sec.\ref{sec:3}, which regards the
\emph{local} stability of Q-balls\footnote{In the above quoted example
of the B-balls, the use of this theorem shows that only B-balls with
$\omega>R^{-1}$ and $\sg^2(0)< M_{p}^2 \exp{|K|^{-1}}$,
where $|K|=0.1-0.01$ and $M_p$ is the Planck mass, are stable. Since we are interested only  in $\sg\ll M_p$, all interesting B-balls are stable.}. But also for $\omega_0^2>0$ this theorem is useful, as it shows that there can be
also Q-balls with more energy than Q free mesons, but locally stable. At
the quantum level this means that they can only decay through tunneling.
It is however not clear how this should take place.

In sec.\ref{sec:4.1} we introduced the thin-wall approximation from a
somehow unusual point of view - as the consequence of a linearization of
the eq. of motion for the core of large Q-balls. The way we did this
makes clear, for a given model, where the approximation breaks. Another advantage of our method is that it gives the
results obtained in \cite{col85,spec87} in an unified and quantitative way. It also becomes clear that the effective potentials discussed in ref.\cite{kus98,enq98,enq99} in the context of AD baryogenesis, don't allow for thin-walled Q-balls. The reason is that if one \emph{turns off} the B-violating non-renormalization terms these potentials are too flat ($\omega_0^2=0$ and $\sg_0=\infty$), and therefore we can't linearize the eq. of motion as we did in sec.\ref{sec:4.1}.

In what concerns the other limit, $\omega_0^2 \to m^2$, we showed that
the thick-wall approximation is useful only when the potential is of the
form $U_3(\sg)=\meio m^2\sg^2-\alpha\sg^3+\cdots$, or $U_4(\sg)=\meio
m^2\sg^2-\lambda\sg^4+\cdots$ - in the first case thick-walled Q-balls
are stable, while in the second one unstable. For more general
potentials, like $U(\sg)= m^2\sg^2-g\Lambda^{4-n}\sg^n + \cdots$, with
$n\geq 6$, we don't know the thick-wall behaviour. In this we desagree with the authors of ref.\cite{mul00}, which used the thick-wall approximation for such potentials. 

Sec.\ref{sec:4.3} discussed the existence of Q-balls \emph{living} in a false vacuum. It is well known that in the limit of nearly degenerate vacua thin-walled Q-balls larger than a certain radius are unstable and can therefore induce a phase transition~\cite{kus97a}. We shown, for the $U_3(\sg)$ theory, that for a very \emph{deep} true vacuum there still are stable Q-balls in the spectrum, although only thick-walled ones.

Finally, we revisited the proof, made in ref.\cite{kus97}, of the stability of thick-walled Q-balls in the model $U_3(\sg)$. The author, trying to determine the second variation of the energy at \emph{fixed} charge, used the method of lagrangean multipliers, which indeed is adequate only for the \emph{first order} variations. The expression obtained in this way (eq.(18) of \cite{kus97}),
\begin{displaymath}
                     \delta^2 E_Q \overset{\text{ref.}[7]}{=} \intx \meio \delta\sg (-\nabla^2 + U''+3\omega^2)\delta\sg,
\end{displaymath}
misses the important \emph{non-local} term that we found in eq.\eqref{cinco} and can be shown to be larger than our result for all values of $\omega$. For instance, with the above expression one would get the result that the translational modes, $\partial_i\sg$, are not zero-modes. As we said above, with our expression we confirm that thick-walled Q-balls are stable in the model $U_3(\sg)$.
 
The expression of ref.\cite{kus97} was later used in ref.\cite{axe00}, for potentials of the form $U_4(\sg)$ in $1+1$-dimensions, as a basis for a numerical calculation of the vibration spectrum, and for the determination of the parameter regions of stable Q-balls and unstable Q-balls. To verify the validity of their results the authors observed numerically the evolution of a Q-ball belonging to the unstable parameter region, and saw that it \emph{realy decays} into plane waves. This positive test, however, is not a convincing argument. The reason is that this Q-ball was picked up \emph{exactly} from the boarding line that separates the \emph{real} unstable region and the \emph{supposed} unstable region in parameter space: they chose a point in parameter space which lies on the line which separates \emph{unbroken} from \emph{broken} symmetry and observed a thin-walled Q-ball which, as we have seen in sec.\ref{sec:4.3}, is an unstable configuration for $\omega_0^2<0$. Would one have picked up a configuration lying in the middle of the \emph{supposed} unstable region, one wouldn't have observed any instability.

Our intention in sec.\ref{sec:5} (and in the appendix) was to show how
the classical Q-balls are related with the properties of the quantum
theory. Usually classical stable configurations are supposed to appear
in the quantum theory as dominant contributions to some functional
integral, since they are extrema of the action or the energy. In this
spirit we did an investigation of a functional integral which furnishes
the lowest lying energy state for a given charge $q$. Although the Q-ball
\emph{does not maximize} the integrand in the original integral, since
the integration runs over all possible configurations, we were able,
after some \emph{cosmetics}, to rewrite the integral in such a way that the
integration runs only over configurations of charge $q$ and the Q-ball of
charge $q$ maximizes the integrand. But, there was a \emph{collateral}
effect - an oscillating term appeared which can erase the contribution
of the neighborhood of the Q-ball to the integral. As we remarked there,
we must compare the period of the oscillation and the thickness of the
gaussian around the Q-ball configuration, to see whether it makes sense to perform a semi-classical expansion or not. In the case of an affirmative answer, the energy of the Q-ball is the zeroth order contribution to the lowest lying energy eigenstate of the given charge and we can use eq.\eqref{eq:76} to calculate radiative corrections to it.

\section*{Aknowledgment}
We would like to thank A.Kusenko for encouraging discussions about Q-balls at the beginning of this work.

\appendix

\section{Functional Integral for fixed charge}

The purpose of this appendix is to derive the expression for $Z_0^q(T)$, eq.\eqref{funcint}, starting from (see eq.\eqref{318}):
\bequ
                    Z_0^q(T)=\int d\phi\, \int_{0}^{2\pi} \frac{d\alpha}{2\pi}\, e^{-i\alpha q}\, \langle \phi|\, e^{i\alpha Q}\, e^{-HT}\, |\phi\rangle.
\end{equation}
A great part of the this derivation follows closely the one by Rajaraman and Weinberg in ref.\cite{raj1}, whose intention was to get to expression \eqref{341}, although in the real time (i.e. Minkowski) formalism.

Our first task is to calculate $\langle \phi|\, e^{i\alpha Q}\, e^{-HT}\, |\phi\rangle$. The effect of applying $e^{i\alpha Q}$ in $\langle\phi|$ is \emph{just} that of changing its phase: $e^{-i\alpha Q}|\phi({\bf x})\rangle = |\phi({\bf x})\,e^{i\alpha}\rangle$. We have thus
\bequ\begin{split}\label{324}
                        \int d\phi \langle \phi|\, & e^{i\alpha Q}\, e^{-HT}\,|\phi\rangle  = \int d\phi \langle \phi\,e^{i\alpha} | \,e^{-HT} \, | \phi\rangle\\ 
& = \int [d\phi] \, \exp{\left(-\int\, dt \,L_E [ \phi]\right)},
\end{split}\end{equation}
where the integration is made over paths with $\phi(T/2)=\phi(-T/2)e^{i\alpha}$.

We now change to the polar coordinates\footnote{Although this change of coordinates is non-trivial, it can be shown that the \emph{naive} substitution of coordinates that we perform is correct up to terms of $4th$ order in the fluctuations around the Q-ball, if we make the change $q^2 \to q^2-1/4$\cite{raj1,pac00}.} $\sg\equiv\sqrt{2}|\phi|$ and $\vartheta\equiv\arg{(\phi)}$ we already used in sec.\ref{sec:3}. With these coordinates the euclidean lagrangian reads
\bequ\begin{split}\label{325}
                        L_E \left[ \sg,\vartheta \right] & = \intx [\tmeio \dot{\sg}^2 + \tmeio (\sg\dot{\vartheta})^2]\\ & \quad + \intx [ \tmeio(\nabla\sg)^2 + \tmeio\sg^2(\nabla\vartheta)^2 + U(\sg)],
\end{split}\end{equation}
which turns to be the energy $E[\sg,\vartheta]$ of the configuration $\sg e^{i\vartheta}/\sqrt{2}$. In the following $E$ will therefore be used in place of $L_E$. Also, the measure is now $[d\phi]=[\sg\,d\sg][d\vartheta]$ and the integration is performed over paths with $\sg({\bf x},T/2) = \sg({\bf x},-T/2)$ and $\vartheta({\bf x},T/2) = \vartheta({\bf x},-T/2) + \alpha$. It is usefull to extend the range of $\vartheta$ from $[0,2\pi]$ to $]-\infty,+\infty[$, a step which poses no problem as its only effect is to multiply the integral with an infinite constant. The same applies to the integration over $\alpha$.

Now, we put the system in a box and expand $\vartheta(x)$ in its Fourier modes:
\bequ\label{326}
                        \vartheta({\bf x},t) = b_{0}(t) + \sum_{{\bf k}_{i}\neq 0}\, b_{{\bf k}_{i}}(t)\, e^{i{\bf k}_{i} {\bf x}_{i}} \equiv b_{0}(t) + \tilde{\theta}({\bf x},t).
\end{equation}
It's easy to recognize that the zero-mode $b_0(t)$ is the only degree of freedom affected by the U(1) internal rotation $\vartheta(x) \to \vartheta(x) +\alpha$. This follows readily from the fact that $\intx \tilde{\theta}({\bf x},t)=0$. The effect of this coordinate transformation in the path integral is thus to change the measure and the integration limits in the following way: $[d\vartheta]=[d\tilde{\theta}]\,[db_{0}]\,J$, $\tilde{\theta}(T/2) = \tilde{\theta}(-T/2)$,
$b_{0}(T/2) = b_{0}(-T/2) + \alpha$ where $J$ is a path independent jacobian. We can now calculate
\bequ
                         Z_0^q = \int_{0}^{2\pi} \frac{d\alpha}{2\pi}\, \int[\sg\,d\sg]\,[d\tilde{\theta}]\,[db_{0}]\,J\, e^{-\int dt\,E - i\alpha q}.
\end{equation}
To do this note that $\alpha =  b_{0}(T/2) - b_{0}(-T/2) = \int dt\, \dot{b}_{0}$, and therefore
\bequ\begin{split}
                         \int\,dt E[\sg & ,\vartheta] + i\alpha q = \int\,dt E[\sg,\tilde{\theta}]+{} \\ 
                                & + \int\,dt \left(\frac{1}{2}\dot{b}_{0}^2\, I[\sg] + \dot{b}_{0}\, \left(Q[\sg,\tilde{\theta}] +iq\right)\right),
\end{split}
\end{equation}
where $I[\sg]$ and $Q[\sg,\tilde{\theta}]$ are the functionals defined in sec.\ref{sec:2}.

Now, we see that the integration in $b_{0}(t)$ combined with the integration in $\alpha$ is the same as integrating over \emph{arbitrary} paths in $b_{0}$, and that the integral is gaussian. Thus, from these integrations we only get a term:
\bequ
                          \left(\prod_{t}\sqrt{I[\sg]}^{-1}\right)\cdot \exp{\left(\int dt \frac{(Q+iq)^2}{2I}\right)},
\end{equation}
in this way obtaining an effective energy~\cite{raj1}
\bequ\label{340}
                          E^{eff}[\sg,\tilde{\theta}] \equiv E[\sg,\tilde{\theta}] - \frac{(Q[\sg,\tilde{\theta}]+iq)^2}{2I[\sg]},
\end{equation}
and the following expression:
\bequ\label{341}
                          Z_0^q(T) = \int[\sqrt{I(\sg)}^{-1}\,\sg\,d\sg]\,[d\tilde{\theta}]\,J\, e^{-\int dt\,E^{eff}},
\end{equation}
where, as we said, $\tilde{\theta}(x)$ includes all non-static spatial Fou\-rier modes, and the paths are cyclical. The energy $E^{eff}$ contains an imaginary part $\text{Im}\,E^{eff} = -qQ[\sg,\tilde{\theta}]/\,I[\sg]$, but this will not contribute to any imaginary part of the path integral, since while Re$\,E^{eff}$ is an even function of $\tilde{\theta}$, Im$\,E^{eff}$ is an odd one and $Z_0^q(T)$ becomes
\bequ\begin{split}\label{eq:87}
                          Z_0^q = \int & [\sqrt{I(\sg)}^{-1}\,\sg\,d\sg]\, [d\tilde{\theta}]\,J \cdot\\
                                  & \quad \cdot e^{-\int dt\,E_q}\cos{\left(q\int dt\,\frac{ Q[\sg,\tilde{\theta}] }{I[\sg]} \right)},
\end{split}\end{equation}
with
\bequ
                          E_q[\sg,\tilde{\theta}]\equiv \text{Re}\,E^{eff}=E[\sg,\tilde{\theta}]+\frac{q^2-Q^2[\sg,\tilde{\theta}]}{2I[\sg]}.
\end{equation}
We will now introduce the following identity under the integral:
\bequ
                       C=\int [db_0(t)]\,\delta\left(\dot{b}-\frac{q-Q[\sg,\tilde{\theta}]}{I}\right),
\end{equation}
where we integrate over all possible paths, and $C$ is an infinite constant, which as usual will be absorbed in the measure. The advantage of this step is that one recovers the integral over all the degrees of freedom and therefore we can rewrite the measure in \emph{cartesian} coordinates. To get rid of $[\sqrt{I}^{-1}]=\prod_{t}(I(t))^{-1/2}$ we absorb it in the $\delta$-function as follows
\bequ
                       I^{-1/2}\delta\left(\dot{b}-\frac{q-Q[\sg,\tilde{\theta}]}{I}\right)=\delta\left(\frac{q-Q[\sg,\vartheta]}{\sqrt{I}}\right),
\end{equation}
where we used $Q[\sg,\vartheta]=Q[\sg,\tilde{\theta}]+\dot{b}I[\sg]$. Since the $\delta$-function imposes $q=Q[\sg,\vartheta]$ we can readily show that $E_q[\sg,\tilde{\theta}]=E[\sg,\vartheta]$:
\bequ\begin{split}
                       E[\sg,\vartheta] & =E[\sg,\tilde{\theta}]+\meio\dot{b}(Q[\sg,\tilde{\theta}]+Q[\sg,\vartheta] \\
                                        & =E[\sg,\tilde{\theta}]+\frac{Q^2[\sg,\vartheta]-Q^2[\sg,\tilde{\theta}]}{2I}.
\end{split}\end{equation}
Finally with the definitions $\phi=\sg e^{i\vartheta}/\sqrt{2}$ and $\alpha=\int dt\,\dot{b}$ we obtain
\bequ\label{eq:fin}
                       \int dt\,\frac{Q[\sg,\tilde{\theta}]}{I[\sg]}=\int dt\,\frac{Q[\phi]}{I[|\phi|]}-\alpha.
\end{equation}
Using equations \eqref{eq:87} to \eqref{eq:fin} we are led to eq.\eqref{funcint}.

%Your text comes here. Separate text sections with
%\section{Section title}
%\label{sec:1}
%and \cite{RefJ}
%\subsection{Subsection title}
%\label{sec:2}
%as required. Don't forget to give each section
%and subsection a unique label (see Sect.~\ref{sec:1}).
%\begin{figure}
%\resizebox{0.75\textwidth}{!}{%
%  \includegraphics{leer.eps}
%\vspace{5cm}       % Give the correct figure height in cm
%\end{figure}
%\begin{figure*}
%\vspace*{5cm}       % Give the correct figure height in cm
%\end{figure*}
%\begin{table}
%\begin{tabular}{lll}
%\hline\noalign{\smallskip}
%\end{tabular}
% Or use
%\vspace*{5cm}  % with the correct table height
%\end{table}
%
% BibTeX users please use
% \bibliographystyle{}
% \bibliography{}

\begin{thebibliography}{}

\bibitem{aff85} I.Affleck, M.Dine, Nucl.Phys. {\bf B249}, (1985) 361.

\bibitem{din96} M.Dine, L.Randall and S.Thomas, Nucl.Phys. {\bf B458}, (1996)
291, hep-ph/9507453.
 
\bibitem{kus98} A.Kusenko, M.Shaposhnikov, Phys.Lett. {\bf B418}, (1998) 46,
hep-ph/9709492.
 
\bibitem{enq99} K.Enqvist, J.McDonald, Nucl.Phys. {\bf B538}, (1999) 321,
hep-ph/9803380.
 
\bibitem{col85} S.Coleman, Nucl.Phys. {\bf B262}, (1985) 263-283.

\bibitem{lee91} T.D.Lee, Y.Pang, Phys.Rep. {\bf 221}, (1992).

\bibitem{kus97} A.Kusenko, Phys.Lett. {\bf B404}, (1997) 285, hep-th/9704073.

\bibitem{pac00} F.Paccetti Correia, Diploma thesis (Heidelberg 2000).

\bibitem{col77a} S.Coleman, Phys.Rev. {\bf D15}, (1977) 2929.

\bibitem{col77b} S.Coleman, Phys.Rev. {\bf D16}, (1977) 1762.

\bibitem{col78} S.Coleman, V.Glaser, A.Martin, Commun.Math.Phys. {\bf 58},
(1978) 211.
 
\bibitem{col87} S.Coleman, Nucl.Phys. {\bf B298}, (1987) 178-186.

\bibitem{lee76} R.Friedberg, T.D.Lee, A.Sirlin, Phys.Rev. {\bf D13}, (1976)
2739. 
\bibitem{spec87} D.Spector, Phys.Lett. {\bf B194}, (1987) 103.

\bibitem{mul00} T.Multam\" aki, I.Vilja, Nucl.Phys. {\bf B574}, (2000) 130,
hep-ph/9908446.
 
\bibitem{kus97a} A.Kusenko, Phys.Lett. {\bf B406}, (1997) 26, hep-ph/9705361.

\bibitem{raj1} R.Rajaraman, E.Weinberg, Phys.Rev. {\bf D11}, (1975) 2950.

\bibitem{raj} R.Rajaraman, {\sl Solitons and Instantons} (North-holland,
Amsterdam, 1982). 
\bibitem{ben89} K.M.Benson, Nucl.Phys. {\bf B327}, (1989) 649.

\bibitem{cal75} C.G.Callan, D.J.Gross, Nucl.Phys. {\bf B93}, (1975) 29.

%\bibitem{ben91} K.M.Benson, L.M.Widrow, Nucl.Phys. {\bf B353}, (1991) 187.

%\bibitem{coh86} A.Cohen, S.Coleman, H.Georgi, A.Manohar, Nucl.Phys. {\bf B272}, (1986) 301.

\bibitem{enq98} K.Enqvist, J.McDonald, Phys.Lett. {\bf B425}, (1998) 309,
hep-ph/9711514.
 
\bibitem{axe00} M.Axenides, S.Komineas, L.Perivolaropoulos, M.Floratos,
Phys.Rev. {\bf D61}, (2000) 085006,
hep-ph/9910388. 
%\bibitem{enq00a} K.Enqvist, J.McDonald, Nucl.Phys. {\bf B570}, (2000) 407.

%\bibitem{enq00b} K.Enqvist, A.Jokinen, J.McDonald, Phys.Lett. {\bf B483}, (2000) 191.

%\bibitem{kas00a} S.Kasuya, M.Kawasaki, Phys.Rev. {\bf D61}, (2000) 041301.

%\bibitem{kas00} S.Kasuya, M.Kawasaki, Phys.Rev. {\bf D62}, (2000) 023512.

%\bibitem{kol90} E.Kolb, M.Turner, {\sl The Early Universe} (Addison-Wesley, 1990 Reading MA).

%\bibitem{kuz85} V.Kuzmin, V.Rubakov, M.Shaposhnikov, Phys.Lett. {\bf B155}, (1985) 36.

%\bibitem{lut95} M.Luty, W.Taylor IV, Phys.Rev. {\bf D53}, (1996) 3399.

%\bibitem{mar97} S.Martin. e-Print Archive: hep-ph/{\bf 9709356}.

%\bibitem{saf88} A.Safian, S.Coleman, M.Axenides, Nucl.Phys. {\bf B297}, (1988) 498.

%Author, \textit{Book title} (Publisher, place year) page numbers
% etc
\end{thebibliography}

\end{document}